\documentstyle[12pt]{article}
\def\copi{\cop^{(int)}}
\def\U{{\cal U}}
\def\hA{\hat A}
\def\hg{\hat g}
\def\dsp{\displaystyle}

\def\hx{\hat x}
\def\hp{\hat p}
\def\beq{\begin{eqnarray}}
\def\eeq{\end{eqnarray}}
\def\lbl{\label}
\def\dx{\dot x}
\def\tR{\tilde R}
\def\dR{\dot R}
\def\hS{\hat S}
\def\1{^{(1)}}
\def\2{^{(2)}}
\def\part#1#2{\frac{\partial #1}{\partial #2}}

\def\om{\omega}

\def\0{^{(0)}}
\def\1{^{(1)}}

\def\e{\epsilon}

\def\({\left(}
\def\){\right)}
\def\[{\left[}
\def\]{\right]}

\def\ben{\begin{enumerate}}
\def\een{\end{enumerate}}
\def\bel{\begin{equation}\label}
\def\ee{\end{equation}}
\def\ba{\begin{array}}
\def\ea{\end{array}}
\def\bt{\begin{tabular}}
\def\et{\end{tabular}}

\def\r#1{(\ref{#1})}
\def\cite#1{[#1]}

\def\tens{\otimes}
\newcounter{popnr}

\def\theequation{\thesection.\arabic{equation}}
\renewcommand{\theequation}{\arabic{section}.\arabic{equation}}
\newcommand{\alpheqn}{\setcounter{popnr}{\value{equation}}
                      \addtocounter{popnr}{1}
                      \setcounter{equation}{0}
   \renewcommand{\theequation}{\arabic{section}.\arabic{popnr}\alph{equation}}}
\newcommand{\reseteqn}{\setcounter{equation}{\value{popnr}}
     \renewcommand{\theequation}
     {\arabic{section}.\arabic{equation}}}
\def\bl{\alpheqn}
\def\el{\reseteqn}
\def\sec{\setcounter{equation}{0}}

\def\~{\widetilde}
\def\cop{\Delta}

\def\to{\,\rightarrow\,}

\def\k{\kappa}
\def\poin{Poincar\'e }
\begin{document}
\begin{titlepage}
\def\thepage{}
\thispagestyle{empty}
\title{Quantum Deformations of Space-Time Symmetries and Interactions}
\author{Jerzy Lukierski \thanks{Supported by KBN grant 2P 302 08706.} \\
\normalsize Institute of Theoretical Physics, University of Wroc{\l}aw,\\
\normalsize pl.\ Maxa Borna 9, 50-203 Wroc{\l}aw, Poland\\[3mm]
Peter C.\ Stichel\thanks{Supported by Alexander von Humboldt-Stiftung.}\\
\normalsize Faculty of Physics, University of Bielefeld,\\
\normalsize Universit\"atsstr.~25, 33615 Bielefeld, Germany
}
\date{}
\maketitle
\begin{abstract}
We discuss quantum deformations of Lie algebra as described by the
noncoassociative
modification of its coalgebra structure. We consider for simplicity the
quantum $D=1$ Galilei algebra with four generators: energy $H$, boost $B$,
momentum $P$ and central generator $M$  (mass generator). We describe
the nonprimitive coproducts for $H$ and $B$ and show that their
noncocommutative and noncoassociative structure is determined by the
two-body interaction terms. Further
we consider the case of physical Galilei symmetry in three dimensions.
Finally we discuss the noninteraction theorem for manifestly covariant
two-body systems in the framework of quantum deformations of $D=4$ \poin
algebra and a possible way out.
\end{abstract}
\end{titlepage}
\def\thepage{\arabic{page}}

\sec
\section{Introduction}
In standard formulation of symmetry schemes the infinitesimal symmetries are
described by classical Lie algebras $\hg$. If we consider the classical
Lie algebra $\hg$ as a bialgebra, we should provide additional information
about the form of the coproduct $\cop(\hg)$. Usually in classical and quantum
mechanics we represent $\hg$
in terms of one-particle observables forming the algebra $\hA$, and
the coproduct $\cop(\hg) \subset \hA \tens \hA$ is a function of
two-particle observables. In particular if $\hg$ describes the group of
motions which includes the energy (generator of time translations) the
choice
\bel{1.1}
\cop(\hg)= \hg\tens1 + 1 \tens \hg
\ee
means that the two-particle system is a free one, because all the space-time
symmetry generators are additive.

One of the aims of this paper is to show that in the category of
noncoassociative bialgebras
which describe space-time symmetries the information
about the form of the coproducts for space-time symmetry generators
carries the information about the two-particle interactions. Because due
to Drinfeld theorem [1] in the formalism of quantum deformations of Lie
algebras it does exist the isomorphism between the
classical $(\U(\hg))$ and deformed $(\U_q(\hg))$ enveloping algebras, it
is possible to transfer all the quantum deformation to the coalgebra
sector (see e.g.\ [2]). Our discussion deals with such a case when the
Lie algebra is classical, but the coproduct (which we recall is a
homomorphism of $\hg$) is nonadditive and noncocommutative. We shall
search for such coproducts as determined by the symmetry-invariant
interactions.
Our result here consists in writing the deformed coproduct for the
generators of space-time symmetries as follows
\bel{1.2}
\cop(\hg) = \cop\0(\hg) + \copi(\hg)
\ee
and relate $\copi(\hg)$ with two-body interactions.

For simplicity we
consider as an example the $D=1$ Galilei algebra. In such a case the
algebra $A$ is the one-dimensional Heisenberg algebra with two
generators $\hx$, $\hp$, and the nonrelativistic space-time algebra
takes the form ($\hg=(B,P,H,M)$)
\bl
\beq
\lbl{1.3a} [B,P]=iM\,, \\{}
\lbl{1.3b} [B,H]=iP\,, \\{}
\lbl{1.3c} [P,H]=[M,\hg]=0\,,
\eeq
\el
where $B$ denotes nonrelativistic boost in two-dimensional space-time
$(x,t)$, $P$ is the momentum (generator of space translations), $H$ is
the energy (generator of time translations), and $M$ is the central
charge generator describing the mass operator. The Galilei symmetry is
described by the following $D=2$ space-time transformations:
\bel{1.4}
\ba{rcl}
x'&=& x+a+vt\,,\\
t'&=& t+b\,.
\ea
\ee
If $M$ takes the numerical mass value $m$, the one-particle realization of the
algebra \r{1.3a}-\r{1.3c} in terms of the canonical variables ($\hx$, $\hp$)
can be chosen as follows (see e.g.\ [3]):
\bel{1.5}
B=m\hx\,,\qquad P=\hp\,,\qquad H=\frac{\hp^2}{2m}\,,\qquad M=m\cdot {
1}\,.
\ee
We see that the relation \r{1.3a} can be identified with the Heisenberg
relation $[\hx,\hp]=i$ (we put $\hbar=1$).

In the following paragraph we shall determine the coproduct of
$\hg=(B,P,H,M)$ by considering the $D=1$ nonrelativistic
Galilei-invariant interacting two-body systems. It should be stressed
that our coproducts are not coassociative.
We shall comment further on the
$D=3$ Galilean case. In the next paragraph we shall discuss from the point of
view of the deformed coproducts the
relativistic \poin invariant two-body systems. In final section we
present the conclusion: that in the Hopf-algebraic (or quantum group)
framework one can understand why large class of Galilei-invariant
interactions are allowed and why there is valid a no-go theorem concerning
four-dimensional covariant two-body interactions.

\sec
\section{The coproduct as the characterization of the two-body
interactions}

Let us consider the following two-body Lagrangian describing Galilei-invariant
interacting two-particle system in one-dimensional space of the
following form $(m>\kappa)$.
\bel{2.1}
L=\frac12 m(\dx^2_1+\dx^2_2)+\frac12\k (\dx^2_1-\dx^2_2)-\frac{\om}2
(x_1-x_2)^2\,.
\ee
It is easy to see that the canonical momenta ($i=1,2$)
\bel{2.2}
p_i =\part{L}{ \dx_i}=m_i \dx_i\,,
\ee
where $m_1=m+\k$, \, $m_2=m-\k$\,.

The energy operator $H$ takes the form
\bel{2.3}
H=\frac{p_1^2}{2m_1}+\frac{p_2^2}{2m_2} + \frac{w}2 (x_1-x_2)^2\,,
\ee
where for $i=1,2$
\bel{2.4}
[\hx_i,\hp_j]= i\delta_{ij}\,.
\ee

The lagrangian \r{2.1} transforms under the special Galilei
transformation ($v\neq0$ in \r{1.4}) as follows
\bel{2.5}
L \to L' =L+\frac{d}{dt}F\,,
\ee
where ($R=\frac{x_1+x_2}2$, $x=x_1-x_2$)
\bel{2.6}
F=2mRv+mv^2t+(c+\k x)v\,.
\ee
The generator of special Galilei transformations is given by the
formula
\bel{2.7}
B=\left.\part{F}{v}\right|_{v=0 \atop t=0} = 2mR +(c+\k x) = 2 m \tR +c\,,
\ee
where $\tR=(2m)^{-1}(m_1x_1+m_2x_2)$. If we define
\bel{2.8}
P=p_1+p_2\,,
\ee
one can check that three generators \r{2.3}, \r{2.7}-\r{2.8} satisfy the
algebra \r{1.3a}-\r{1.3b}, whith $M=2m$.

Below we shall write these generators in the coproduct form.
We shall consider both $\k$ and $\om$ in the Lagrangean \r{2.1} as
describing the interaction terms.
Because
(see \r{1.5}; $i=1,2$):
\bel{2.9}
x_i =\frac1m B_i\,,\qquad P_i=p_i\,,\qquad H_i=\frac{p_i^2}{2m}\,,\qquad M_i=m\,,
\ee
one can write the two-body generators in the following form (we choose
$c=0$, or shift $B\to B-c$):
\bl
\beq
\lbl{2.10a} B&=& \frac{m_1}m B_1 + \frac{m_2}m B_2\,,\\
\lbl{2.10b} H&=& \frac{m}{m_1}H_1+\frac{m}{m_2}H_2 + \frac{\om}{2m^2}
(B_1-B_2)^2\,,\\
\lbl{2.10c} P&=& P_1+P_2\,,\qquad M=M_1+M_2\,,
\eeq
\el
or using the language of coproduct we should write
\bl
\beq
\lbl{2.11a}\cop(B)&=& \frac{m_1}m B\tens 1 +\frac{m_2}m 1\tens B\,,\\
\lbl{2.11b}\cop(H)&=& \frac{m}{m_1} H\tens 1 +\frac{m}{m_2} H\tens 1\\
\nonumber && +\frac\om{2m^2} (B^2\tens1 - 2 B\tens B + 1 \tens B^2)\,,\\
\lbl{2.11c}\cop(P)&=& P\tens 1 +1 \tens P\,,\qquad \cop(M) = M\tens 1 + 1 \tens M\,.
\eeq
\el
It is easy to see that the coproducts \r{2.11a}-\r{2.11c} are the
homomorphism of classical $D=1$ Galilei algebra provided the central
mass generator $M$ is diagonal, i.e. $M=m\cdot { 1}$.

We would like to make the following comments:
\ben
\item[i)] The space-dependent harmonic potential modifies the primitive
coproducts for the energy generator. Here we show that if the interaction
modifies kinematic term by introducing the split of masses  ($\k\neq0$),
then we obtain the nonadditivity of the boost generators.
\item[ii)]
It appears that the coproduct for $B$ can be made quite complicated by
replacing in \r{2.1} the ``mass-difference'' kinematic term by the
following general velocity-dependent interaction
\footnote{Such a term has been discussed very recently by J.\
\L{}opusza\'nski and P.\ C.\ Stichel [9].}
\bel{2.12}
\frac{\k}2 (\dx_1^2-\dx_2^2) \to \k(x) \dx_i \dR_i,,
\ee
where $\dx_i\dR_i=\frac12(\dx_1^2-\dx_2^2)$. We see that the choice
$\k(x)=\k$ corresponds to the case considered above. For the choice of
general function $\k(x)$ one obtaines after the substitution \r{2.12}
the formulae \r{2.3} with $m_1\to m_1(x)=m+\k(x)$, \, $m_2\to m_2(x)=
m-\k(x)$. Because in the coalgebra language one can write
\bel{2.13}
x=x_1-x_2 \equiv \frac B m \tens 1 - 1 \tens \frac B m\,,
\ee
the formula \r{2.10a} after the substitution \r{2.12} can be
generalized in the straightforward way.
Further the harmonic potential can be replaced in \r{2.1}
with general velocity-dependent potential
\bel{2.14}
\frac{\om^2}2 x^2 \to U(x,\dx)
\ee
The modification \r{2.14} of \r{2.1} affects only the Hamiltonian \r{2.3},
in which the potential energy is replaced with momentum-dependent
potential
\bel{2.15}
H= \frac{p_1^2}{2m_1}+\frac{p_2^2}{2m_2}+ V(x,p)\,,
\ee
where $p=(2m)^{-1}(m_2p_1-m_1p_2)$. Using the substitution \r{2.14} and
\bel{2.16a}
p \to \frac1{2m}(m_2\cdot P\tens 1 - m_1 \cdot { 1} \tens P)
\ee
 one can rewrite \r{2.15} as the nontrivial coproduct
for $H$.
\item[iii)]
In all considered cases, also after the substitutions \r{2.12} and \r{2.14},
the momentum variable remains additive, i.e.\ the primitive coproduct \r{2.10a}
for $P$ remains valid in the presence of arbitrary Galilei-invariant interactions.
\item[iv)]
It is interesting that for the coproducts \r{2.11a}--\r{2.11c} the
coassociativity condition
\bel{2.16n}
(\cop\tens 1) \cop(\hg) = (1 \tens \cop) \cop(\hg) \,,\qquad \hg=H,B,P,M
\ee
is \underline{not} valid for any choice of the two-body potential
$V(x,p)$ (in particular this statement can be checked for the choice of
coproducts
\r{2.11a}--\r{2.11c}). It appears that the coassociativity imposes very
stringent restrictions on the class of allowed interactions which if are
possible should have very unconventional form. It should be mentioned
that the violation of coassociative structure in quantum mechanics has been
recently discussed in [10].
\item[v)]
The present discussion can be also extended to the case of $D=3$
Galilean system. The realization \r{1.5} is generalized in
straightforward way with $\hx$ and $\hp$ replaced by the three-vectors,
and quantum-mechanical realization $M_i=\e_{ijk}\hx_j \hp_k$ for
the angular-momenta. For the case with spin the one-particle realization
of $D=3$ Galilei algebra takes the form
\bel{2.18}
\ba{rcl}
M_i&=&\dsp \e_{ijk}\hx_j\hp_k+\hS_i\,,\qquad B_i=m\hx_i\,\\[3mm]
P_i&=&\dsp p_i\,,\qquad H_i=\frac{\hp_i \hp_i}{2m}\,,\qquad M_i=m\,,
\ea
\ee
i.e.\ the 8-dimensional phase space ($\hx_i$, $\hp_i$, $\hS_i$), where
$S_i^2$ is a Casimir, can be reexpressed in terms of $D=3$ Galilei
generators. In particular any interaction $U(\vec x,\vec p,\vec x \vec S,
\vec p \vec S)$ (where $\vec x$ is relative coordinate, $\vec p$ is relative
momentum and $\vec S$ is the total spin)
in the Hamiltonian can be described by the correction to the
primitive coproduct for  the energy operator $H$.
\een

\sec
\section{Relativistic interactions and no-interaction theorem}

It is well-known that in comparison with \poin invariance
the Galilean invariance is much less restrictive
for the construction of invariant two-body interactions.
There are possible the following two approaches to
relativistic Poincar\'e-invariant interactions:
\ben
\item[i)] Let us consider for simplicity spinless relativistic particles.
One assumes that the relativistic system is described by
eight-dimensional phase-space ($\hx_\mu$, $\hp_\mu$) where
\bel{3.1}
[\hx_\mu,\hp_\nu]=i\eta_{\mu\nu}\,,
\ee
with some additional constraints imposed. In such a case e.g.\  two-body potentials
would be described by the function $V(x_\mu)$, where
$x_\mu=x_\mu\2-x_\mu\1$. Only the space coordinates
$x_\mu$ ($\mu=1,2$) can be expressed with the \poin algebra generators
$M_i$, $N_i$, $P_i$, $P_0$ (where $M_i$ describe space rotations, $N_i$
-- three relativistic boosts and $P_\mu = (P_i,P_0)$ -- the fourmomenta) by
the following formula
\bel{3.2}
X_i=P_0^{-\frac12}N_i P_0^{-\frac12}\,,
\ee
Because the two-body covariant potential $V(x_\mu)$ depends also on relative
time variable, it can not be reexpressed in
terms of a pair of \poin generators describing two relativistic
particles. Concluding it is not possible to describe the energy of
covariantly interacting two-particle system as a coproduct of the \poin
algebra generator $P_0$.
\item[ii)]
One can also impose the conditions  of relativistic invariance by
considering only the three-dimensional coordinates and momenta
$x_i^{(k)}$, $p_i^{(k)}$ (for two-body systems $i=1,2$). The \poin
generators describing relativistic interacting 2-particle system should
satisfy the following conditions [3-6]:
\bel{3.3}
\ba{rcl}
[M_i,x_j^{(k)}]&=& i\e_{ijk}x_j^{(k)}\,,\\[2mm]
[L_i,x_j^{(k)}]&=& ix_i^{(k)} [H,x^{(k)}]
\ea
\ee
with
\bel{3.4}
P_i=p_i\1+p_i\2
\ee
Let us assume that the two-body \poin symmetry generators satisfy the
classical \poin algebra. In such a framework the problem of the existence
of covariant
interaction has been studied very extensively [4-6]. Under the assumption
that the Hamiltonian is of standard type:
\bel{3.5}
\det \left|\frac{\partial^2 H}{\partial p_i^{(k)}\partial p_j^{(l)}} \right| \neq
0\,,
\ee
it has been shown [4] that the two-body \poin generators are the sum of
two free-particle realizations:
\bel{3.6}
\ba{l}
M_i=\frac12 \e_{ijk}x_jp_k +S_i\,,\qquad P_i=p_i\,,\qquad P_0=(\vec p^2
+m^2)^{\frac12}\,,\\[2mm]
N_i=(\vec p^2 + m^2)^{\frac12} x_i + [m+(\vec p^2 +m^2)^{\frac12}
]^{-\frac12}\e_{ijk}p_j S_k\,,
\ea
\ee
i.e.\ the manifestly covariant relativistic two-body
interaction is not allowed.

We see therefore that the relativistic two-body interaction (under some technical
assumptions see \r{3.6}) is not allowed. In the language of quantum deformations of
the \poin algebra it appears that there does not exist a nontrivial
coproduct for classical \poin algebra generators satisfying the
condition \r{3.4}, i.e.
\bel{3.7}
\cop(P_i) = P_i \tens 1 + 1 \tens P_i\,.
\ee
Indeed, the same conclusion is obtained from the consideration of the
$\k$-deformed \poin algebra [7, 8] in the classical algebra basis [2]. It
follows from these considerations that the coproduct relation \r{3.7}
has to be modified.
\een
\section{Final Remarks}

The aim of this paper is to relate the symmetry-invariant interactions
with the existence of the deformation of the coproducts for classical
symmetry algebras. It appears that if we do not require coassociativity
the nonrelativistic Galilean symmetry
is not so restrictive --- indeed it is possible to find very large class
of the deformed coproducts satisfying classical Galilei algebra
relations. Quite different situations occurs for relativistic systems.
In the framework of standard manifestly covariant relativistic interaction
satisfying relations \r{3.4}-\r{3.5} the two-body relativistic potential
is not allowed. However we know (see e.g. [2]) that there do
exist the deformations of coproducts for classical \poin algebra, with
nonadditive three-momentum generators i.e.\ with the relation
\r{3.7} modified. We conjecture that such deformed Poincar\'e-Hopf
algebras describe permitted class of the relativistic interactions, however
in new category of nonstandard Hamiltonians.
These Hamiltonians will have nonlocal structure in time,
with modified kinetic terms containing derivatives of
arbitrarily high order.
The explicit forms of such
nonstandard manifestly covariant relativistic interactions are now
under consideration.

\subsection*{Acknowledgement}
One of the authors (J.L.) would like to thank Anatol Nowicki for the
discussion of the noncoassociativity problem.
\newpage
\def\b#1{\item}
\def\jmp#1{{\em Journ. Math. Phys. }{\bf#1}}
\def\jp#1{{\em Journ. Phys. }{\bf#1}}
\def\pl#1{{\em Phys. Lett. }{\bf#1}}
\def\mpl#1{{\em Mod. Phys. Lett. }{\bf#1}}
\def\rmp#1{{\em Rev. Mod. Phys. }{\bf#1}}
\def\ncim#1{{\em Nuovo Cim. }{\bf#1}}

\end{document}